\documentstyle[sprocl]{article}

\input{psfig}
\def\Journal#1#2#3#4{{#1} {\bf #2}, #3 (#4)}

\def\NPB{{\em Nucl. Phys.} B}
\def\PLB{{\em Phys. Lett.}  B}
\def\PRL{\em Phys. Rev. Lett.}

\def\be{\begin{equation}}
\def\ee{\end{equation}}
\def\bea{\begin{eqnarray}}
\def\eea{\end{eqnarray}}
\def\Tr{{\rm Tr}}

\begin{document}

\title{SU(5)+ADJOINT HIGGS AT FINITE TEMPERATURE}

\author{ A. RAJANTIE }

\address{Department of Physics, P.O. Box 9, FIN-00014 University of 
Helsinki, Finland}

%%%%%%%%%%%%%%%%%%%%%%%%%%%%%%%%%%%%%%%%%%%%%%%%%%%%%%%%%%%%%%
% You may repeat \author \address as often as necessary      %
%%%%%%%%%%%%%%%%%%%%%%%%%%%%%%%%%%%%%%%%%%%%%%%%%%%%%%%%%%%%%%

\maketitle\abstracts{
I present a three-dimensional effective theory that describes the
high-temperature equilibrium behavior of the SU(5) theory accurately, 
but is much easier to study using non-perturbative methods than the
original four-dimensional one. The effective theory is obtained by 
perturbatively integrating out the super-heavy Matsubara modes as well 
as the heavy temporal component of the gauge field. Regardless of the
particle spectrum of the original theory, the resulting three-dimensional 
theory contains only a gauge field and an adjoint Higgs field. The phase
diagram of the theory is analysed perturbatively and it is shown to have
a cellular structure in three-dimensional parameter space.
}
  
\section{GUT transition}
If the Standard Model is a effective low-energy theory of a grand unified
theory, there may have been a GUT phase transition in the very early
universe. It would have taken place when the temperature was $T\sim 10^{16}$ 
GeV and would have had important cosmological effects.

Topological defects are necessarily created in the transition. 
Depending on the precise
form of the GUT, these may be strings, domain walls or textures,
but in any case there are monopoles, since the fundamental group
of the Standard Model gauge group SU(3)$\times$SU(2)$\times$U(1)
is non-trivial. The defects are important, since they can give rise to
structure formation in the universe. However, the formation of
a high density of monopoles is in conflict with standard cosmology and
with observations. This monopole problem can be solved if the
universe enters an exponentially expanding, inflationary phase
in this or in some other transition taking place soon after the GUT one.
Since GUTs do not conserve baryon number it is also possible that
baryon asymmetry was generated. $B+L$ would be washed out later in the
electroweak transition, but $B-L$ would remain. However, the
simplest GUT candidate SU(5) does not violate $B-L$.

\section{SU(5) dimensional reduction}
From the low-energy physics one cannot deduce the form of the GUT.
To understand the consequences of the transition even
on a qualitative level, one still needs to have some concrete theory
with which to make the calculations. The obvious choice is the
SU(5) model suggested by Georgi and Glashow\cite{ref:Georgi}, since
it has the simplest structure. Furthermore, we drop the fermion
fields and the fundamental Higgs field, since they are inessential
when describing the transition. Thus we have a theory with a
gauge field $A_\mu$ and an adjoint Higgs field $\Phi$. The Lagrangian
of the theory is
\be
\label{eq:orig4d}
{\cal L}_M=-\frac{1}{2}\Tr F_{\mu\nu}F^{\mu\nu}
+\Tr D_\mu\Phi D^\mu\Phi
-m^2\Tr\Phi^2
-\lambda_1(\Tr\Phi^2)^2
-\lambda_2\Tr\Phi^4,
\ee
where $D_\mu\Phi=\partial_\mu\Phi+i g[A_\mu,\Phi].$

In the perturbative broken phase the system can be parametrized by
the vector mass $M$, the SU(3) octet scalar mass $m_8$, the
neutral scalar mass $m_1$ and the gauge coupling constant that
is chosen to have the value $g\approx 0.39$ given by the running of
the Standard Model coupling constants.

Since a gauge symmetry cannot be broken, the only non-perturbatively
meaningful operators are gauge-invariant.
Thus the simplest local fields are
\be
{\rm Tr}\Phi^n,\quad {\rm Tr}F_{\mu\nu}\Phi^{(n-1)}, \mbox{ where }
2\le n\le 5.
\ee
In the broken phase these can be identified with the photon and
a neutral scalar. In the symmetric phase all fields are composite.

To investigate the transition we start from the standard finite-temperature
formalism. The temporal dimension gets replaced by a compact one, which
can be integrated out perturbatively in dimensional 
reduction\cite{ref:dimred}.
No infrared divergences appear in this calculation since they 
arise from the static modes which are still present 
in the effective three-dimensional
theory. The calculation is performed by comparing one- and two-loop
Green's functions in the 4d finite-T theory and the 3d effective theory.
Since the temporal component of the gauge field $A_0$ gets a heavy
Debye mass, it can be also be integrated out. Thus we are left with
an effective theory of the form
\be
\label{equ:effth}
{\cal L}=\frac{1}{2}\Tr F_{ij}F_{ij}
+\Tr(D_i\overline\Phi)^2
+\overline{m}^2\Tr\overline\Phi^2
+\overline\lambda_1(\Tr\overline\Phi^2)^2
+\overline\lambda_2\Tr\overline\Phi^4.
\ee
%where the relation between the 3d and 4d parameters is
%\bea
%\overline\Phi^2
%&=&\frac{1}{T}
%\left[1-\frac{g^2}{16\pi^2}15L_b\right]\Phi^2\nn
%\overline A_i^2
%&=&\frac{1}{T}
%\left[1
%+\frac{g}{24\pi}\sqrt{\frac{5}{2}}
%-\frac{g^2}{16\pi^2}\left(10L_b+\frac{5}{3}\right)\right]A_i^2\nn
%\overline g^2
%&=&
%g^2T\left[
%1-\frac{g}{24\pi}\sqrt{\frac{5}{2}}+\frac{g^2}{16\pi^2}
%\left(\frac{35}{2}L_b+\frac{5}{3}\right)\right]\nn
%\overline\lambda_1
%&=&
%T\left\{\lambda_1-\frac{3g^3}{8\sqrt{10}\pi}\right.\nn
%&&\left.
%-\frac{1}{16\pi^2}
%\left[
%\left(
%32\lambda_1^2+\frac{94}{5}\lambda_1\lambda_2+\frac{84}{25}\lambda_2^2
%+\frac{9}{2}g^4-30g^2\lambda_1\right)L_b-3g^4\right]\right\}\nn
%\overline\lambda_2
%&=&T\left\{\lambda_2-\frac{g^3}{8\pi}\sqrt\frac{5}{2}\right.\nn
%&&\left.
%-\frac{1}{16\pi^2}
%\left[
%\left(
%12\lambda_1\lambda_2+\frac{32}{5}\lambda_2^2
%+\frac{15}{2}g^4-30g^2\lambda_2\right)L_b-5g^4\right]\right\}\nn
%\overline m^2
%&=&m_3^2-\frac{5g^3T^2}{4\pi}\sqrt\frac{5}{2}
%+\frac{g^4T^2}{16\pi^2}
%\left(
%\frac{75}{2}\log\frac{m_3+2m_D}{2m_D}-\frac{25}{4}\right)\nonumber
%\eea
The details of the calculation and the relations between
4d and 3d parameters are given in Ref.\cite{ref:OmaSU5}.

The form of the effective potential does not depend on
the original particle spectrum. All the fermions are integrated out
since they have no static modes and all the other scalars get an
effective mass from the thermal corrections and are integrated out
along with the $A_0$ field. Thus their only contribution is to change
the relation between the 3d and 4d parameters. 
This justifies the assumption that they are inessential in this
context.

\section{Effective theory}
The parameters of the 
effective theory (\ref{equ:effth}) are dimensionful. Thus we can use one 
of them  to fix the scale and parametrize the theory
by
three dimensionless numbers
\be
y=\frac{\overline{m}^2(\overline{g}^2)}{\overline{g}^4},\quad
x_1=\frac{\overline{\lambda}_1}{\overline{g}^2},\quad
x_2=\frac{\overline{\lambda}_2}{\overline{g}^2}.
\ee
The theory is superrenormalizable and only the mass parameter must be
renormalized.
Its running can be calculated exactly in a two-loop perturbative
calculation.

The relation between 4d and 3d parameters given in terms of the perturbative
broken phase masses is
\bea
x_1&\approx&0.21\frac{m_1^2}{M^2}-0.59\frac{m_8^2}{M^2}-0.012,\quad
x_2\approx2.52\frac{m_8^2}{M^2}-0.020,\nonumber\\
y&\approx&6.80+6.08\frac{m_8^2}{M^2}-1.45\frac{m_8^4}{M^4}
+3.95\frac{m_1^2}{M^2}-0.038\frac{m_1^4}{M^4}\nonumber\\
&&-0.12\frac{m_8^2m_1^2}{M^4}
-21.9\frac{m_1^2}{T^2},
\eea
where the logarithmic divergence on the temperature is neglected.
Thus, as temperature decreases, $x_1$ and $x_2$ stay constant and
$y$ decreases. Thus $y$ can be interpreted as a measure of
temperature.

\begin{figure}
\centerline{\psfig{file=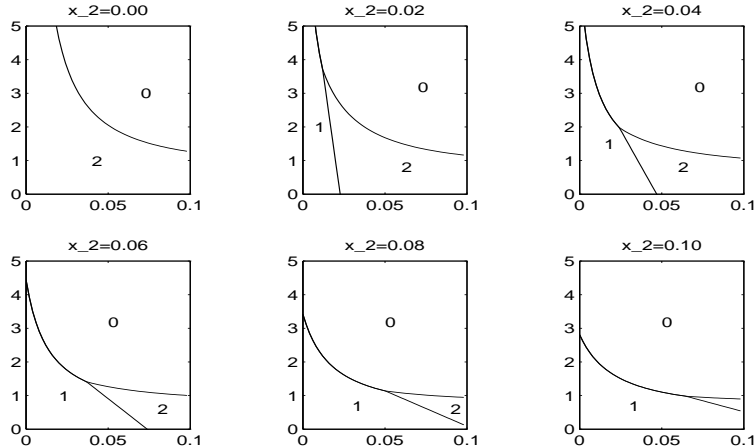,height=6cm,width=10cm}}
\caption{Cross-sections of the perturbative phase diagram 
of the effective theory at various values of $x_2$. 
Horizontal and vertical axes corresponds to $x_1$ and $y$, respectively.
0=symmetric phase, 1=SU(3)$\times$SU(2)$\times$U(1), 2=SU(4)$\times$U(1).}
\label{fig:phases}
\end{figure}

The effective theory can be analysed in perturbation theory. There
are three different phases,
\be
\langle\Phi\rangle=0,\quad
\langle\Phi\rangle\sim{\rm Diag}(2,2,2,-3,-3),\quad
\langle\Phi\rangle\sim{\rm Diag}(1,1,1,1,-4).
\ee
Calculation of the two-loop effective potential in both of the broken 
phases gives the phase diagram  shown if Fig.~\ref{fig:phases}.
As the universe cools down, the SU(5) symmetry is broken down to
either SU(4)$\times$U(1) or SU(3)$\times$SU(2)$\times$U(1). The transitions
between the phases are always first-order ones in perturbation theory.
However, lattice simulations on analogous, but simpler 
models\cite{ref:SU2}
suggest that at large values of $x_1$ and $x_2$ there is a
crossover instead of a transition. This is possible since there
is no symmetry breakdown connected to this transition.

A numerical simulation would be the most reliable way to investigate the
transition. On a lattice the theory (\ref{equ:effth}) obtains the form
\bea
{\cal L}_{\rm latt}
&=&
\frac{1}{a^4g^2}\sum_{i,j}\Tr\left[{\bf 1}-P_{ij}(x)\right]\nonumber\\
&&+
\frac{2}{a^2}\sum_i
\left[\Tr\Phi(x)^2-\Tr\Phi(x)U_i(x)\Phi(x+i)U_i^{-1}(x)\right]\nonumber\\
&&+
m^2\Tr\Phi^2+\lambda_1(\Tr\Phi^2)^2+
\lambda_2\Tr\Phi^4,
\eea
where $a$ is the lattice spacing. Owing to superrenormalizability the
lattice parameters can be related to the continuum ones with a
two-loop calculation in lattice perturbation theory. In practice
this means calculating the two-loop effective potential.
This has been performed explicitly in Ref.\cite{ref:OmaLatt}.
This gives a connection between the 4d continuum and 3d lattice 
parameters.

At present there are no lattice simulations of this model.
They would be interesting also independently of the original GUT
model, since SU(5) has much more structure than e.g.~SU(2).
Thus its phase transition could exhibit some interesting 
non-perturbative phenomena.

\section{Conclusions}
We constructed explicitly a three-dimensional purely bosonic
effective theory that describes the equilibrium
behavior of the SU(5) GUT near the transition. We also analysed 
it perturbatively and found its phase diagram. However, lattice
simulation is needed to get reliable information. To this end we
calculated the relation between lattice and continuum parameters.
However, no simulations have yet been made. 

In analogous but simpler models the transition is a crossover for
some parameter values. If the same is true here, it would change
the picture of the GUT transition. If there is only a crossover,
the cosmological consequences are much smaller than in the
first-order case, since the system can stay near the equilibrium.
In practice this rules out both baryogenesis and inflation in
this transition, but
does not cure the monopole problem, since the monopoles formed
by thermal fluctuations still fall out of equilibrium and their
density remains far too high.

\section*{Acknowledgments}
I am grateful to the Alfred Kordelin Foundation for financial support.

\end{document}